\documentstyle[aps,prl,twocolumn]{revtex}
\input epsf
\def \kpi  {K/\pi}
\def \kpip {K^+/\pi^+}
\def \kpim {K^-/\pi^-}
\def \kpkm {K^+/K^-}
\def \np   {N_p}
\begin{document}
\draft

\wideabs{
\title{Systematic Study of the Kaon to Pion Multiplicity Ratios 
	in Heavy-Ion Collisions}

\author{F.~Wang$^{1}$, H.~Liu$^{3,1}$, H.~Sorge$^{2}$, N.~Xu$^{1}$, 
	and J.~Yang$^{4,1}$}

\address{
$^1$ Nuclear Science Division, Lawrence Berkeley National Laboratory, 
	Berkeley, California 94720\\
$^2$ Department of Physics, State University of New York at Stony Brook, 
	Stony Brook, New York 11794\\
$^3$ Physics Department, Kent State University, Kent, Ohio 44242\\
$^4$  Physics Department, University of California at Los Angeles, 
	Los Angeles, California 90095
}

\maketitle

\bigskip
\begin{abstract}
We present a systematic study of the kaon to pion multiplicity ratios 
($\kpip$ and $\kpim$) in heavy-ion collisions from AGS to RHIC energy 
using the Relativistic Quantum Molecular Dynamics (RQMD) model.
The model reasonably well describes the available experimental
data on $\kpip$ and $\kpim$.
Within the model, we find that the strong increase of the ratios 
with the number of participants is mainly due to hadronic rescattering
of produced mesons with ingoing baryons and their resonances.
The enhancement of $\kpi$ in heavy-ion collisions with
respect to elementary $p+p$ interactions is larger at AGS 
than SPS energy, and decreases smoothly with bombarding energy.
The total multiplicity ratios at RHIC energy are predicted 
by RQMD to be $\kpip=0.19$ and $\kpim=0.15$.
\end{abstract}

}

\section{Introduction}\label{introduction}

Nuclear matter at high energy density has been extensively 
studied through high energy heavy-ion collisions~\cite{qm,harris}. 
The primary goal of these studies is to observe the possible 
phase transition from hadronic matter to quark-gluon plasma 
(QGP)~\cite{qgp,McL86}, in which quarks and gluons are deconfined 
from individual hadrons forming an extended region. 
The phase transition is predicted by lattice QCD calculations 
to occur at a temperature 140--170~MeV and an energy density
on the order of 0.5--1.5~GeV/fm$^3$~\cite{lattice}.
It is believed that the QGP state existed in the early universe 
shortly after the big bang~\cite{McL86},
and may also exist in the cores of neutron stars~\cite{star}.

If a QGP is produced in a heavy-ion collision, the collision
system will evolve in stages from deconfined quarks and gluons
to interacting hadrons, and finally to freeze-out particles
which are detected. In order to extract the information
about the postulated quark-gluon plasma stage of heavy-ion collisions,
systematic studies of multi-observables at freeze-out as a function
of the collision volume and bombarding energy are necessary.
These observables include strangeness production~\cite{strangeness,Egg91},
charm production~\cite{charm}, lepton production~\cite{lepton},
jet quenching~\cite{jet_quench}, 
elliptic~\cite{elflow}, and other types of flow~\cite{flow}.
For a critical review of these observables, see Ref.~\cite{Bas99}.

In this article, we constrain ourselves to one of 
the above observables, namely, strangeness production. 
In particular, we focus on the kaon to pion multiplicity ratios
($\kpi$ for both charge signs), since the bulk of strangeness 
produced in heavy-ion collisions are carried by kaons.

It is not clear how much the $\kpi$ ratios in heavy-ion reactions 
reflect the properties of the phase transition between QGP and 
hadronic matter. 
Originally it was argued that $\kpi$ may serve as a signature 
of the QGP, and indeed early measurements at the BNL AGS and 
CERN SPS~\cite{early_data} showed a significant enhancement 
of $\kpip$ in heavy-ion collisions over $p+p$ interactions. 
The idea was that, in the deconfined state, strange quark pairs 
($s\bar{s}$) may be copiously produced through gluon-gluon fusion
($gg \rightarrow s\bar{s}$)~\cite{Egg91,Kap86}, 
while in the hadronic gas such pairs are produced via 
pairs of strange hadrons at a higher production threshold. 
Therefore the time needed for a hadronic gas system in kinetic 
equilibrium to reach chemical equilibration is significantly 
longer than the life time of a heavy-ion collision which is 
typically on the order of 10~fm \cite{Egg91,Kap86,Lee88}.
This idea, however, has been challenged because it neglects 
pre-equilibrium dynamics of the initial stage which may 
considerably speed up chemical equilibration~\cite{Mat89}:
Initial interactions between produced particles and ingoing baryons 
are ``harder'' than in kinetic equilibrium. 
This is confirmed by detailed transport calculations 
which can reproduce the early experimental data on $\kpip$ 
\cite{Mat89,Sor91:meson,Sor95:mass,Sa93}.

Even on the level of equilibrium physics, it is not entirely clear
which difference to expect in a comparison of the scenarios with and 
without phase transition. 
Strangeness in a chemically equilibrated hadronic gas might be 
as high as or higher than in a QGP \cite{Kap86,Lee88,McL87}. 
In fact, thermal hadron gas model requires a so-called strangeness 
suppression factor to fit to experimental particle ratios;
this has been interpreted as some indication of non-equilibrium 
transition from QGP to hadron gas~\cite{Gaz}.
In this picture, a ``lower'' than expected $\kpi$ ratio 
is a signature of the QGP.

Recently several measurements of $\kpi$ have been made at both the AGS 
and SPS by different experiments \cite{Wan96,Ahl98,E866,NA44,NA49}.
There has been renewed interest in the interpretation
of the measured $\kpi$ ratios~\cite{Gaz,Cassing}.
Interestingly, the Giessen group found the strongest deviations of 
their hadronic model (HSD) calculations from experimental data not 
at SPS energy but at the lower AGS energy~\cite{Cassing}.

Despite of the theoretical mist, the $\kpi$ ratios may still be valuable 
observables to be studied not only to address questions of the phase 
transition but also to obtain a better understanding of 
the pre-equilibrium dynamics, the hadronization processes, 
and dynamics of hadrons in the medium.
In these studies, the beam energy is an essential control parameter 
whose variation allows to modify two important variables: 
the initial baryon and energy densities. 
Both densities are of importance as to whether the system enters into 
the quark-gluon phase or remains hadronic all the time. 
If the ``strangeness content'' of the hadronic matter is substantially
different from a QGP, discontinuity might be expected in the 
excitation function of the $\kpi$ ratios. 
With data from RHIC emerging in the near future, where QGP is 
likely to be formed, the first things to look for are changes in the 
observables departing from what have been observed at low energies.

In this paper we present a systematic study of the $\kpi$ ratios 
using the Relativistic Quantum Molecular Dynamics (RQMD) model 
\cite{rqmd,Sor92:rope,Sor95:flavor}.
For simplicity, we only consider the equal and
large mass nucleus-nucleus collisions. 
The goal of this paper is to provide an understanding of the 
underlying mechanisms for $\kpi$ enhancement by comparing model 
results with available experimental data at various energies. 
We also present predictions of the $\kpi$ ratios at RHIC energy.
Throughout the paper, $\kpi$ stands for the integrated
total multiplicity ratios unless otherwise notified.

The paper is organized as follows. In Sec.~\ref{sec_rqmd},
we give brief description of the RQMD model. 
In Sec.~\ref{sec_pp}, we demonstrate that the calculated 
pion and kaon multiplicities are in a good agreement 
with experimental data in elementary $p+p$ interactions.
In Sec.~\ref{sec_aa}, we present our results on $\kpi$ 
ratios in calculated heavy-ion collisions.
This section is divided into four parts. 
Part~A presents a systematic study of $\kpi$ as a function of 
the number of participants,
parts~B and C investigate the effects of particle rescattering 
and rope formation on the ratios, 
part~D considers excitation functions of $\kpi$ spanning AGS, SPS, and 
RHIC energies ($\sqrt{s}\approx 5$, 17 and 200~AGeV, respectively), 
and Part~E derives the $\kpi$ enhancement factors.
Section~V summarizes our findings.

\section{The RQMD model}\label{sec_rqmd}

RQMD is a semi-classical microscopic model which combines
classical propagation with stochastic interactions~\cite{rqmd}.
The degree of freedom in RQMD depends on the relevant length and 
time scales of the processes considered.
In low energy collisions ($\sim$1~AGeV), RQMD reduces to solving 
transport equations for a system of nucleons, other hadrons
and resonances interacting in binary collisions or via mean field. 
At high beam energies ($^>_{\sim}$10~AGeV), color strings and 
hadronic resonances are excited in elementary collisions; 
their fragmentation and decay lead to production of particles.
Overlapping strings do not fragment independently but form 
``ropes'' \cite{Sor95:mass,Sor92:rope,Sor95:flavor,Bir84}. 
The secondaries which emerge from the fragmenting strings, 
ropes and resonances may re-interact. 
For all the results presented here, the so-called 
``cascade mode'' is used in RQMD (no ``mean field''). 

In order to calculate heavy-ion collisions at RHIC energy, 
the RQMD code evolved from version 2.3 to 2.4. 
The physics are the same in the two versions. 
The change of version was due to technical reasons.
We use version 2.3 in calculations for all results 
presented in this paper, except for heavy-ion collisions 
at RHIC energy version 2.4 is used.
We have checked our AGS and SPS results with those from 
version 2.4 and our RHIC results of peripheral heavy-ion collisions
with those from version 2.3; consistencies were found.

Let us note about the ``rope'' mechanism implemented in the model.
In RQMD \cite{Sor95:mass,Sor92:rope,Sor95:flavor}
(and also in QGSM~\cite{QGSM} and the Spanish version of DPM~\cite{DPM}),
strings fuse into color ``ropes'' if they overlap in transverse and
longitudinal configuration space; the transverse dimension of 
strings is on the order of 0.8~fm~\cite{Sor95:mass}.
The strong $s\bar{s}$ enhancement in the coherent rope fields is
a consequence of the large chromoelectric field strength, 
because quark pair production rates depend strongly on the ratio
between the squared quark mass and the field strength~\cite{Won94}.
It has been shown that the rope mechanism strongly enhances 
multi-strange baryon production \cite{Sor95:mass,Sor95:xi,Sor97:NN97}. 
However, its effect on single-strange hadron production is small 
once particle rescattering is considered, 
resulting in negligible change in the $\kpi$ ratios~\cite{Sor95:flavor}.

On the other side, rescattering between hadrons and resonances changes 
considerably the ratio between produced strange and non-strange quarks. 
This change is about 80\% for central Pb+Pb collisions 
at SPS energy~\cite{Sor95:flavor}.
Multi-step processes such as
\begin{tabbing}
\hspace{25mm}
\=$\pi+N$\=$\searrow$\=$\Delta$\=+\=$\varrho$\=$\swarrow$\=$\pi+\pi$\kill\\
\>$\pi+N$\>          \>        \> \>         \>          \>$\pi+\pi$     \\
\>       \>$\searrow$\>        \> \>         \>$\swarrow$\>              \\
\>       \>          \>$\Delta$\>+\>$\varrho$\>          \>              \\
\>       \>      \>\>\hspace{0.6mm}$\downarrow$\>\>      \>              \\
\>       \>          \>  \>$\Delta^\star$\>  \>          \>
\end{tabbing}
are of essential importance for strangeness production, 
because intermediate resonances act as an ``energy store.''
Such multi-body interactions are only frequent in
a system of sufficient density of roughly 1 fm$^{-3}$,
because the lifetime of the intermediate states is
typically 1--2 fm/$c$~\cite{Hof95}. 
Therefore, strangeness enhancement via rescattering is not 
expected to occur in the late dilute stages of the reactions.

Since feed-down affects the extracted particle yields,
let us finally mention which hadrons are kept stable 
when we calculate the $\kpi$ ratios.
Members of the baryon octet (mainly $\Lambda,\Xi,\Omega$), and the 
$K^0_S, K^0_L, \eta$, and $\phi$ mesons are not decayed in the model.
In reality, all these particles decay into pions via 
single and/or multi-step channels; 
$\phi$'s and $\Omega$'s decay into kaons in addition.
Unless notified otherwise, these decay products are not included 
in the charged pion or kaon multiplicity. 
According to RQMD, the $\eta$-decay contribution to the charged 
pion multiplicities is 2--4\%;
$K^0_S$ multiplicity is, to a good approximation, one half of the 
charged kaon multiplicity, therefore the $K^0_S$-decay contribution 
to pion multiplicities can be deduced from $\kpip$ and $\kpim$;
the contributions from all other sources to $\pi^+$ ($\pi^-$) 
multiplicity are less than 3\% (5\%) and 5\% (7--15\%),
respectively for $p+p$ and heavy-ion collisions;
and the $\phi$-decay contributions are a couple to 10\% in $K^+$  
and 7--15\% in $K^-$ multiplicity in all collisions studied.

\section{RQMD results for $p+p$ interactions}\label{sec_pp}

\begin{figure}
\centerline{
\epsfxsize=0.52\textwidth\epsfbox[30 40 480 540]{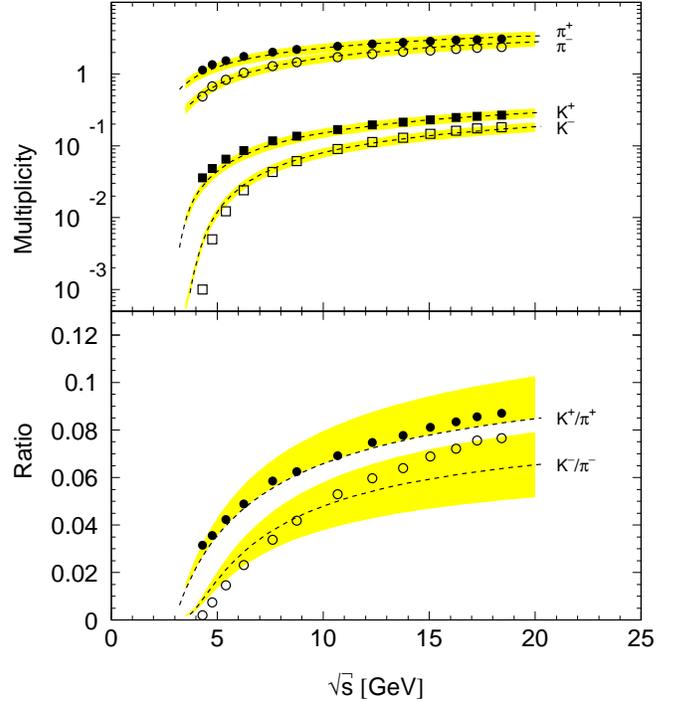}}
\caption{
RQMD results of $p+p$ interactions at various center-of-mass 
energies ($\sqrt{s}$).
Top panel: $\pi^+$ (filled circles), $\pi^-$ (open circles),
$K^+$ (filled squares), and $K^-$ (open squares) total multiplicities.
Bottom panel: $\kpip$ (filled circles) and $\kpim$ (open circles) ratios.
All statistical errors on the RQMD results are smaller than the symbol size.
The dashed curves are parametrizations of experimental 
data~[\protect{\ref{Ros75:pp}}].
The shaded areas around the curves in the top panel reflect 15\% 
experimental uncertainties.
Those in the bottom panel reflect the propagated 21\% uncertainties 
(only one half of the experimental uncertainties on $\kpip$
is shown for clarity).}
\label{fig:pp}
\end{figure}

Figure~\ref{fig:pp} (top panel) shows the inclusive $\pi^+$, $\pi^-$, 
$K^+$, and $K^-$ multiplicities in $p+p$ interactions as a function of 
the center-of-mass energy ($\sqrt{s}$) from AGS to SPS energy. 
The $\kpip$ and $\kpim$ ratios are shown in the bottom panel.
The symbols are results of RQMD calculations.
The dashed curves are parametrizations
of the experimental $p+p$ data~\cite{Ros75:pp},
with experimental uncertainties shown in the shaded areas.
Good agreement is found between the RQMD results and 
the experimental data, especially for $\sqrt{s} > 6$~AGeV.
We note that similar degrees of agreement have also been 
achieved for $p+A$ collisions~\cite{Sor91:meson,Sor93:pA,pA}. 
These agreements provide a realistic base for RQMD calculations 
of heavy-ion collisions.

It should be noted that the agreement between the RQMD results
and the experimental $p+p$ data is not granted automatically by      
inputs into RQMD, such as parametrizations of exclusive cross sections.
Depending on energy, resonances or strings are excited 
in the elementary inelastic nucleon-nucleon collisions 
(see Refs.~\cite{Sor95:flavor,Sor90:string} for details).
While the resonance parameters are taken from the data tables, the
parameters of string fragmentation are fixed from the properties
of strings created in $e^+e^- \rightarrow q\bar{q}$ annihilations. 
One of the most important differences concerning particle
production in $e^+e^-$ versus hadron-hadron interactions is 
the fragmentation of the ingoing valence quarks 
(e.g., the leading-particle effect~\cite{Sor93:pA}).
It is treated in a constituent quark spectator fragmentation 
approach which keeps track of the gluonic ``junction'' 
connecting all three quarks of a nucleon. 
This is important for the stopping of nucleons on a nuclear target.

\section{RQMD results for heavy-ion collisions}\label{sec_aa}

We have calculated Au+Au collisions at lab beam energies
8, 11, 14.6, 20, and 30~AGeV 
($\sqrt{s}=4.3, 4.8, 5.4, 6.3$, and $7.6$~AGeV),
Pb+Pb collisions at lab beam energies
40, 60, 80, 100, 120, 140, 158, and 180~AGeV
($\sqrt{s}=8.8, 10.7, 12.3, 13.8, 15.1, 16.3, 17.3$, and $18.4$~AGeV),
and Au+Au collisions at RHIC energy ($\sqrt{s}=200$~AGeV).
The comparisons of $\kpip$ between RQMD and lower energy data are
reported elsewhere~\cite{Sor97:NN97}.
By studying these collisions, we shall try to identify the underlying 
physics within RQMD for the enhancement in $\kpi$ ratios at the AGS and 
SPS energies, and present RQMD predictions for the ratios at RHIC energy.

\subsection{$\kpi$ systematics versus centrality}

In this section, we study $\kpi$ ratios as a function of the collision
centrality. We choose the number of participants ($\np$) as a
characterization of the collision centrality, considering that $\np$ 
is experimentally accessible via forward energy measurements.
For RQMD results, $\np$ is taken to be the number of initial
nucleons that interact at least once with other particles. 

\begin{figure}[hbt]
\centerline{
\epsfxsize=0.5\textwidth\epsfbox[0 40 500 540]{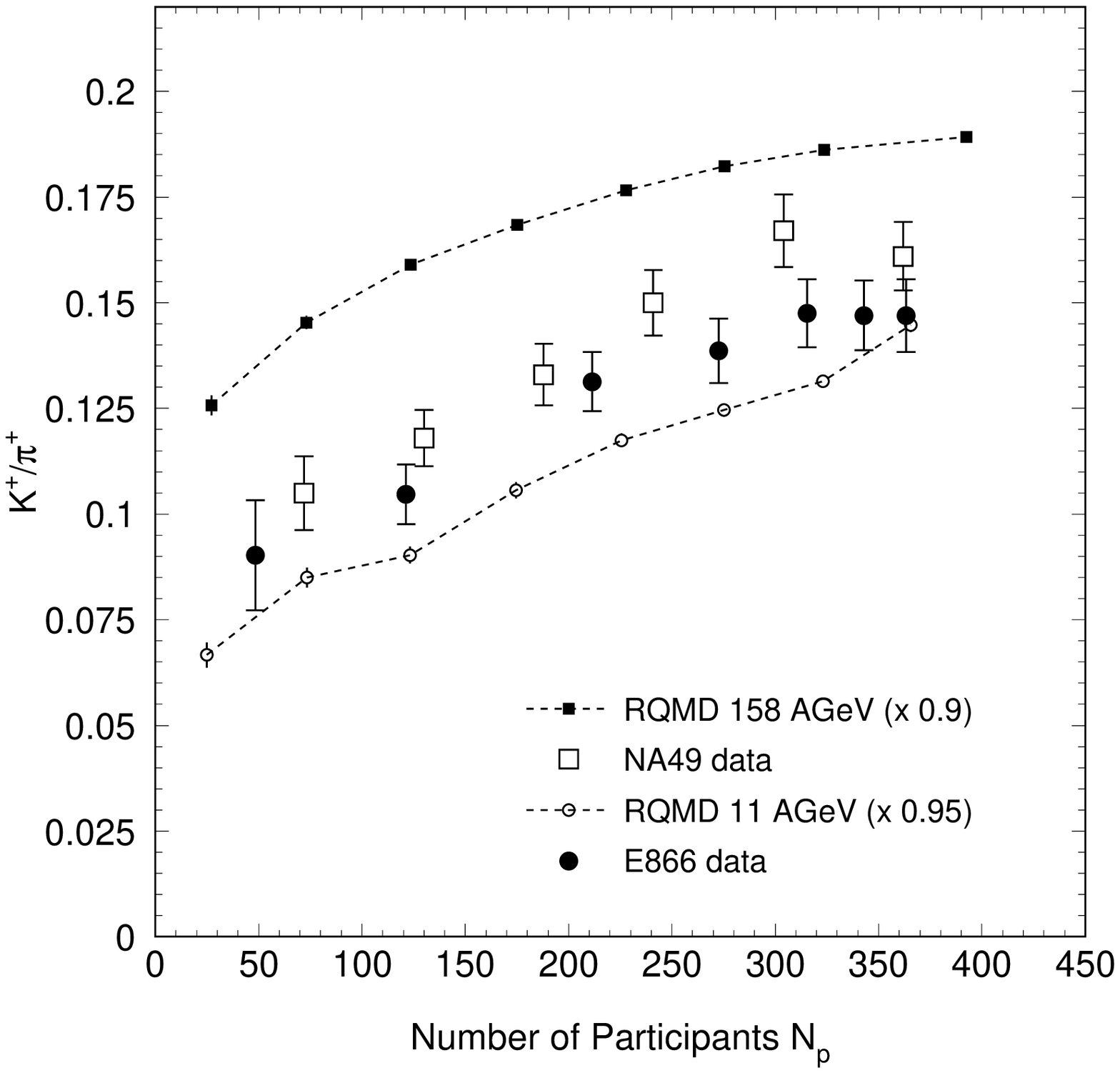}
\epsfxsize=0.5\textwidth\epsfbox[0 40 500 540]{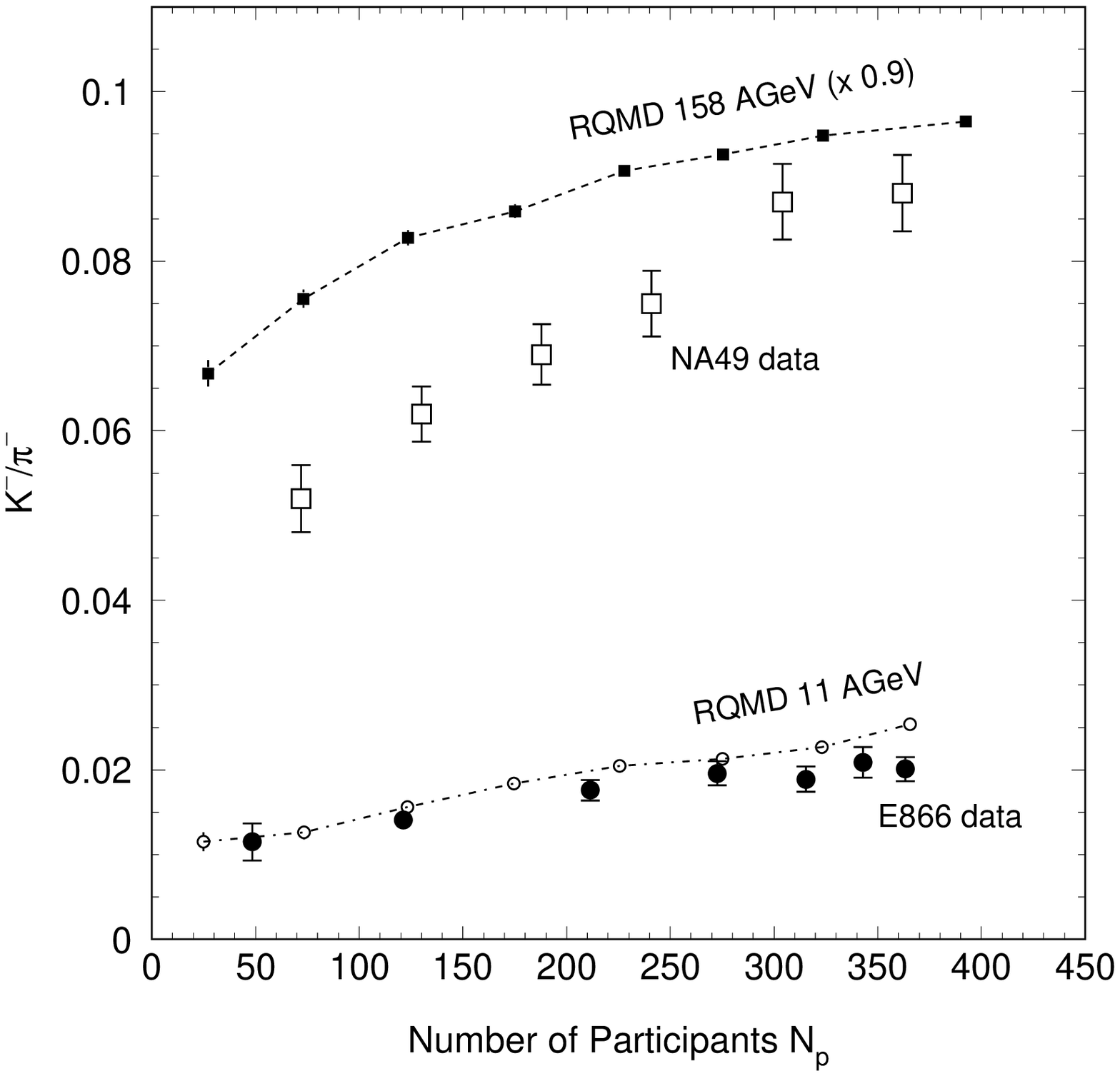}}
\caption{
Comparison of the kaon to pion total multiplicity ratios, 
$\kpip$ (left panel) and $\kpim$ (right panel), between RQMD results 
and preliminary data from Au+Au/Pb+Pb collisions at AGS and SPS energies 
as a function of the number of participants ($\np$).
The RQMD results are shown in the small symbols connected by the lines.
The filled circles are preliminary data from AGS E866~[\protect{\ref{Wan96}}].
The open squares are preliminary data from SPS NA49~[\protect{\ref{NA49}}].
In order to facilitate the comparison, the RQMD results are 
scaled down to include decay contributions.
Errors (either shown or smaller than the symbol size) 
are statistical only for both the RQMD results and the E866 data.
Errors on the NA49 data are quadratic sum of statistical and 
5\% systematic errors.
}
\label{fig:npart_data}
\end{figure}

Preliminary experimental data on the centrality dependence of $\kpi$ 
are available from AGS E866~\cite{Wan96} and SPS NA49~\cite{NA49}. 
Before discussing the RQMD results at various energies, we compare
the RQMD results at AGS and SPS energies to the available data 
in Fig.~\ref{fig:npart_data}. 
In order to facilitate the comparison, contributions from 
$\eta, \phi, K^0_S$, and strange (anti)baryon decays 
are included in the RQMD results.
These contributions reduce the RQMD $\kpip$ ratio 
by 5\% (and is negligible for $\kpim$) at AGS energy,
and by 10\% for both $\kpip$ and $\kpim$ at SPS energy.
Good agreement is found between the RQMD results and the AGS data 
(within 20\%).
RQMD systematically overpredicts the ratios at SPS energy by 10--30\%. 
The agreement is fair with regard to the relatively large systematic 
errors on the data. See Parts~B and C for further discussion.

\begin{figure}[hbt]
\centerline{
\epsfxsize=0.5\textwidth\epsfbox[0 40 500 540]{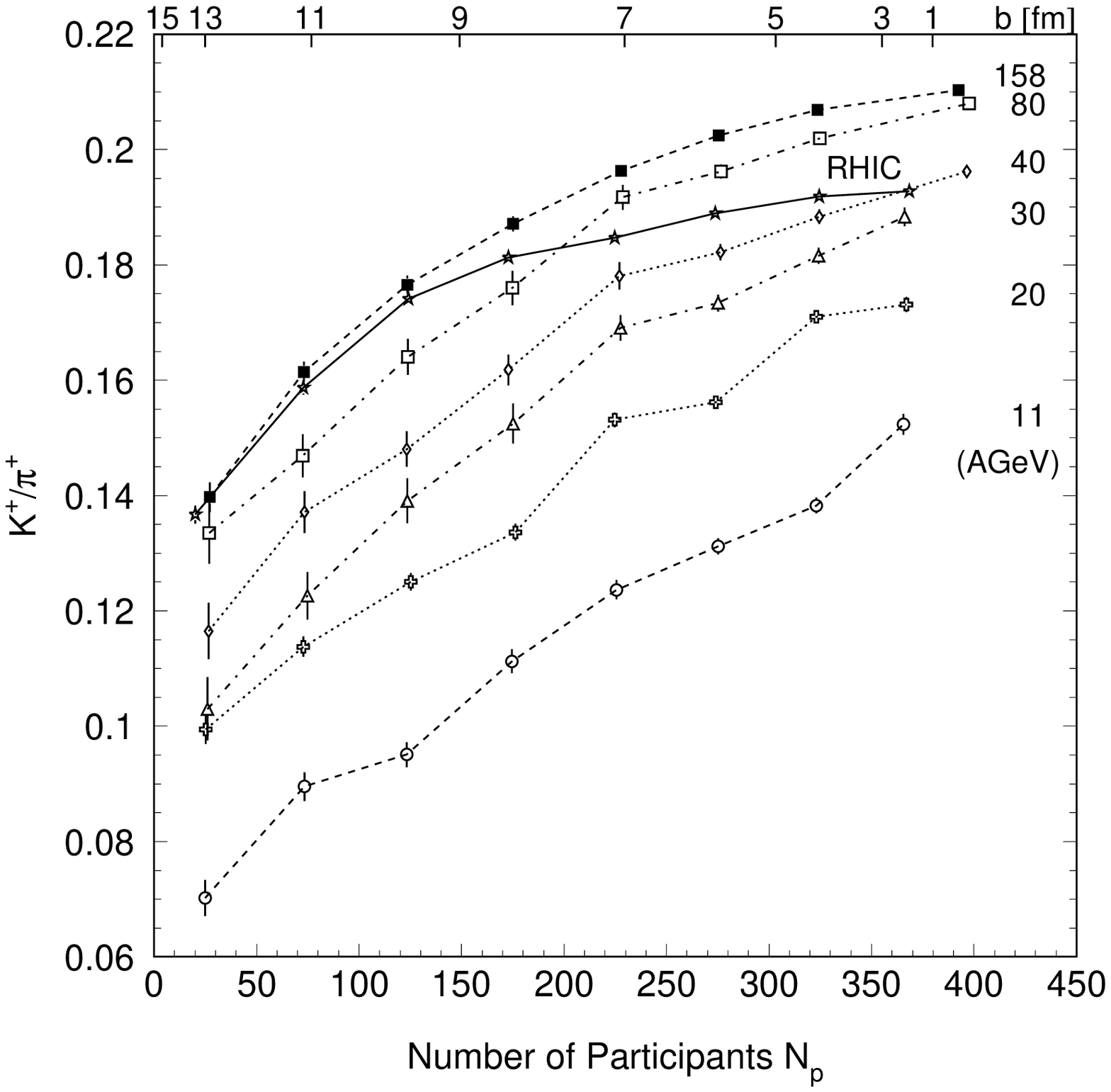}
\epsfxsize=0.5\textwidth\epsfbox[0 40 500 540]{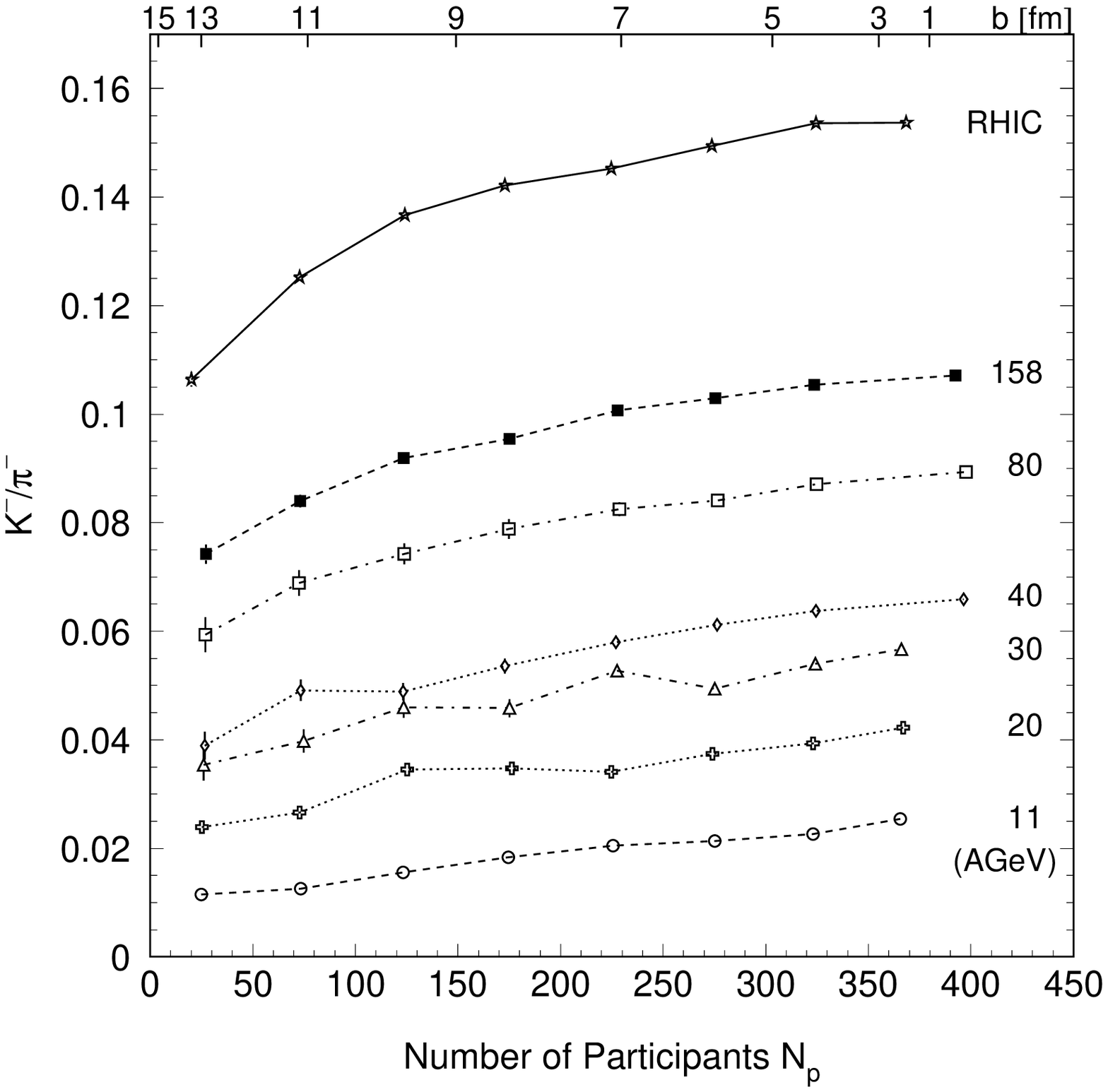}}
\caption{
RQMD results of the kaon to pion total multiplicity ratios,  
$\kpip$ (left panel) and $\kpim$ (right panel), in Au+Au/Pb+Pb 
collisions at various bombarding energies as a function of 
the number of participants ($\np$). 
The lines are labeled by the lab beam energies 
except for RHIC where $\sqrt{s}=200$~AGeV.
Errors are statistical only, either shown or smaller than the symbol size.
The top axis indicates the average impact parameter ($b$) 
for Au+Au collisions at RHIC energy from default RQMD. 
The average impact parameters for other energies 
have a systematic variation of $\pm 0.5$~fm from these values.}
\label{fig:npart}
\end{figure}

Figure~\ref{fig:npart} (left panel) shows $\kpip$ 
calculated by RQMD as a function of $\np$
in heavy-ion collisions at various bombarding energies.
$\kpip$ increases with $\sqrt{s}$ and increases from peripheral 
to central collisions. 
For all collision energies studied except RHIC energy,
the dependence of the ratio on $\np$ is similar.
The increase in the value of $\kpip$ is of the same magnitude from 
peripheral to central collisions.
Consequently, the relative increase in $\kpip$ from peripheral 
to central collisions is larger at lower energies.

Figure~\ref{fig:npart} (right panel) shows $\kpim$ as 
a function of $\np$.
$\kpim$ has similar dependence on $\np$
as $\kpip$ for collisions at energies between AGS and SPS. 
At each centrality, the ratio increases with $\sqrt{s}$.
Unlike $\kpip$, the absolute increase in $\kpim$ 
from peripheral to central collisions is larger at higher energies.

At RHIC energy as shown by Fig.~\ref{fig:npart}, 
the $\np$ dependence of $\kpip$ is distinctly 
different from those at the low AGS and SPS energies. 
Here, a saturation of the ratio seems to appear in central collisions.
Consequently, the absolute increase in the ratio from
peripheral to central collisions is smaller.
On the other hand, $\kpim$ continues to increase with $\np$, 
similar to those observed at the lower energies. 
The absolute value of $\kpim$ at RHIC energy is 
significantly higher than the low energy values.
It is also interesting to note that the absolute value of 
$\kpip$ is lower at RHIC than SPS energy in central collisions. 

What makes $K^+$ and $K^-$  different is, of course, 
the presence of the net baryon number. 
The effect of the fixed net baryon number is larger at lower energies. 
As a result, associate production of $K^+$ via $NN \rightarrow NK^+Y$ 
(where $Y = \Lambda$ or $\Sigma$) dominates over that of $K^-$ 
via $\overline{N}\; \overline{N} \rightarrow \overline{N}K^-\overline{Y}$.
Note that pair production results in the same number of $K^+$'s and $K^-$'s.
On the other hand, the difference between associate production of
$K^+$ and $K^-$ becomes negligible at RHIC energy;
particle production 
from string and rope decays becomes important.
Therefore, $\kpip$ and $\kpim$ display a more similar shape and magnitude 
at RHIC than SPS energies, as shown in Fig.~\ref{fig:npart}. 

\subsection{Effect of particle rescattering}

It is shown in Refs.~\cite{Mat89,Sor91:meson}, within the 
framework of RQMD, that the increase in $\kpip$ in heavy-ion
collisions at AGS energy with respect to $p+p$ and $p+A$ 
is largely due to secondary rescattering among the particles.
Therefore, we shall study the $\kpi$ ratios as a function of the 
average number of collisions ($\nu_p$) suffered by each nucleon.
$\nu_p$ includes both nucleon-baryon and nucleon-meson interactions.
Interestingly, $\nu_p$\ shows little variation with 
bombarding energy; it depends mainly on the collision geometry.

For both $\kpip$ and $\kpim$, the $\nu_p$ dependence 
displays a similar feature seen in Fig.~\ref{fig:npart}.
These similarities are not surprising because the average values 
of $\nu_p$ and $\np$ are correlated (although the distribution
of $\nu_p$ versus $\np$ is fairly broad).
For large $\nu_p$, however, $\kpip$ appears to saturate for 
nuclear collisions at high $\sqrt{s}$ including those in 
the SPS energy regime.

To elaborate further the effect of particle rescattering on 
the $\kpi$ ratios, we calculate the ratios using RQMD with 
particle rescattering turned off.
To be specific, we include only the primary interactions 
of the ingoing nucleons (including rope formation) 
while the projectile and target pass through each other, 
but turn off the meson-induced interactions. 
The results of $\kpip$ and $\kpim$ are shown in the dashed curves 
in Fig.~\ref{fig:npart_ags_sps} for heavy-ion collisions at AGS
and SPS energies. 
The RHIC energy results are shown in Fig.~\ref{fig:npart_rhic}.
For comparison, the default RQMD results are also shown.
The top axis indicates the average impact parameter $b$ corresponding
to $\np$. The $b$ values in Fig.~\ref{fig:npart_ags_sps} are extracted 
from Pb+Pb collisions at SPS energy, and in Fig.~\ref{fig:npart_rhic}
from Au+Au collisions at RHIC energy.
Note that the average $b$ has a systematic deviation of $\pm 0.5$~fm
for collisions at AGS energy and with different settings
(default versus no rescattering and/or no rope).

Referring to Fig.~\ref{fig:npart_ags_sps} and Fig.~\ref{fig:npart_rhic}, 
the $\kpip$ and $\kpim$ ratios obtained with no particle rescattering
show a shallow increase (if not constant) with $\np$ in heavy-ion 
collisions at all three energies.
Thus meson-induced interactions are responsible for
the significant increase in the ratios with $\np$ obtained 
with the default RQMD, especially at AGS energy.
This confirms the early results obtained in Refs.~\cite{Mat89,Sor91:meson}.
Meson-induced interactions mainly increase the kaon production rate; 
they do not change the pion multiplicity significantly.
The changes in the pion multiplicities due to rescattering 
are $-20$\%, $-10$--0\% and $+20$\% at AGS, SPS, and RHIC energies, 
respectively. 
Likewise, the corresponding changes in the kaon multiplicity
are 300\%, 20-30\% and 60\%.
Consequently, the $\kpi$ ratios are increased by including
meson-induced interactions.

The above results can be understood if one keeps in mind that 
resonances play the most important role in order to overcome 
kaon production threshold.
To illustrate this, compare $\pi\Delta (1232)$ and $\pi N$ collisions 
at the same relative momentum 700~MeV/$c$ in the rest frame.
In addition to pion production, the $\pi\Delta$ collision has 
a certain probability to produce a $K\Lambda$ pair.
On the other hand, the $\pi N$ collision can only produce pions.
Thus $\pi N $ collisions do not contribute 
directly to the strangeness enhancement.
Only as a part of a many-body process do they play a role 
(e.g., by involving a $\varrho $ as an intermediate state 
which subsequently interacts with another hadron).
Similarly, the contribution from $\pi \pi $ collisions turns out 
to be irrelevant. 

With the meson-induced interactions turned off, $\nu_p$
is significantly lower than obtained by the default RQMD,
because only the nucleon-baryon interactions are counted.
At AGS energy, the $\kpi$ ratios are found to be constant 
over $\nu_p$, indicating that multiple baryon-baryon 
interactions do not alter the $\kpi$ ratios.
They increase the individual kaon and pion multiplicities 
with similar relative magnitude.
At the higher SPS and RHIC energies, the physics in RQMD is
different from that at AGS energy: Strings and ropes are formed 
and their fragmentation results in particle production.
Thus, the meaning of $\nu_p$ is questionable.

Since the rope mechanism is a novel implementation in RQMD
in an attempt to take into account the string-string interactions
during the hot and dense stage of heavy-ion collisions,
we study its effect on $\kpi$ in the next section. 

\subsection{Effect of rope formation}

\begin{figure}[hbt]
\centerline{
\epsfxsize=0.5\textwidth\epsfbox[0 40 500 540]{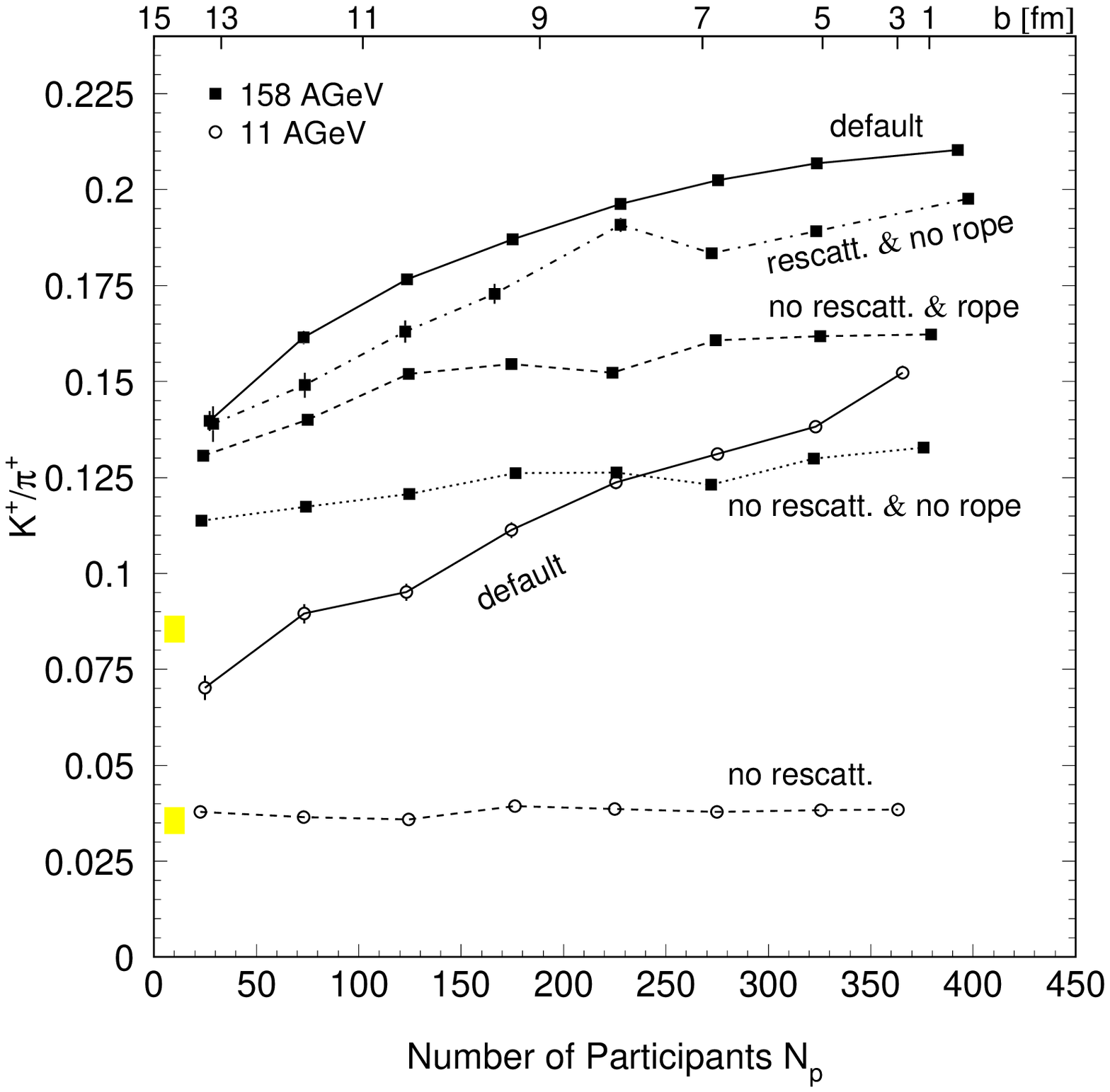}
\epsfxsize=0.5\textwidth\epsfbox[0 40 500 540]{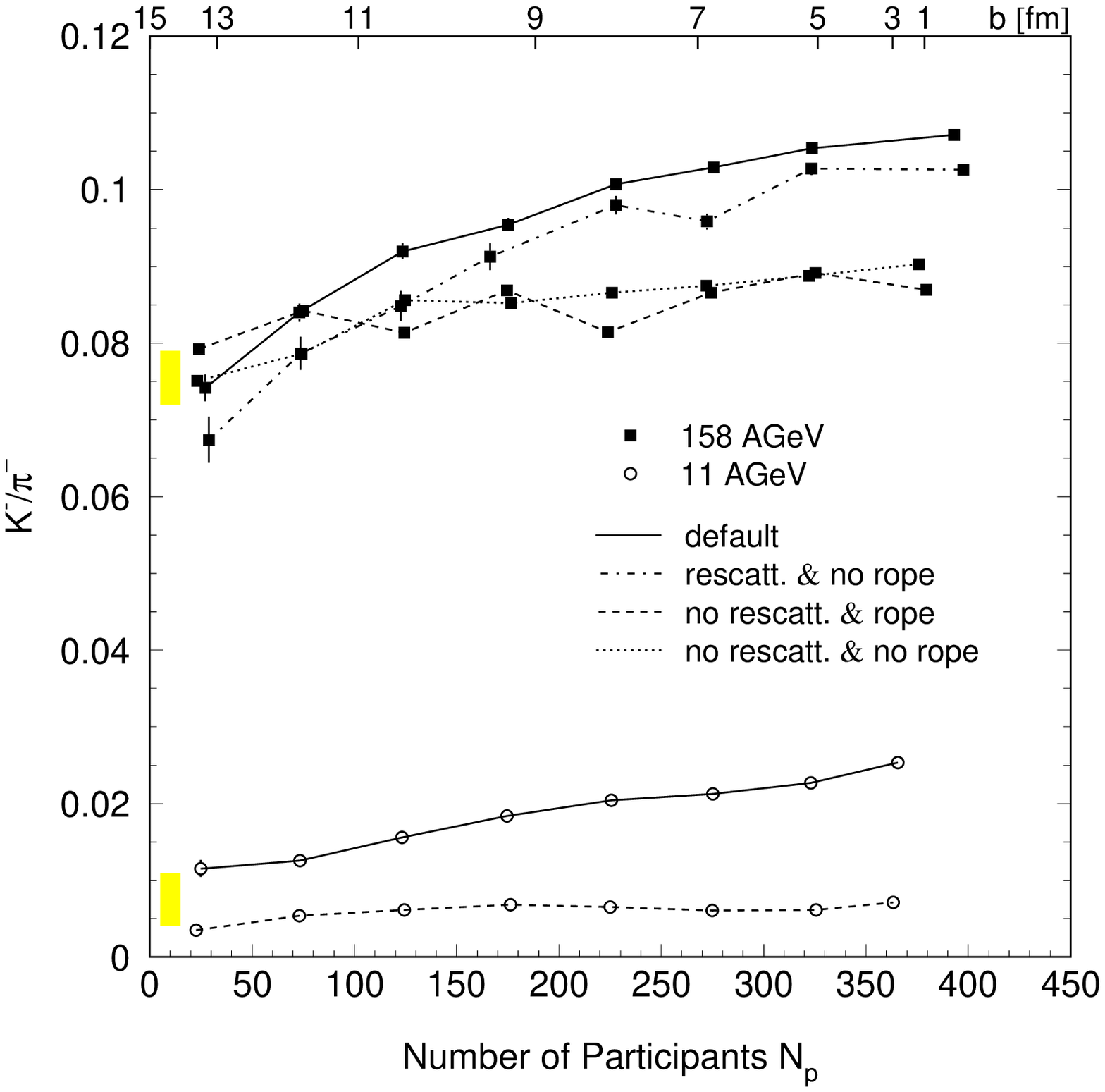}}
\caption{
RQMD results of the kaon to pion total multiplicity ratios, 
$\kpip$ (left panel) and $\kpim$ (right panel), in Au+Au collisions 
at AGS energy and in Pb+Pb collisions at SPS energy as a function of 
the number of participants ($\np$). 
Shown are the RQMD results with meson-induced interactions and rope
formation both turned off (dotted line), with meson-induced interactions
turned off but rope formation included (dashed lines), 
with meson-induced interactions included but rope formation turned off
(dot-dashed line), and with the default setting (solid lines).
The AGS results are shown with the open circles, and 
the SPS results are shown with the solid squares.
Rope formation has negligible effect at AGS energy.
Errors are statistical only, either shown or smaller than the symbol size.
The shaded areas indicate the corresponding ratios in $p+p$ interactions
calculated by default RQMD at AGS (lower) and SPS (upper) energies.
The top axis indicates the average impact parameter ($b$) 
for Pb+Pb collisions at SPS energy from default RQMD. 
The average impact parameters for AGS energy and for
the other settings 
have a systematic variation of $\pm 0.5$~fm from these values.}
\label{fig:npart_ags_sps}
\end{figure}

\begin{figure}[hbt]
\centerline{
\epsfxsize=0.5\textwidth\epsfbox[0 40 500 540]{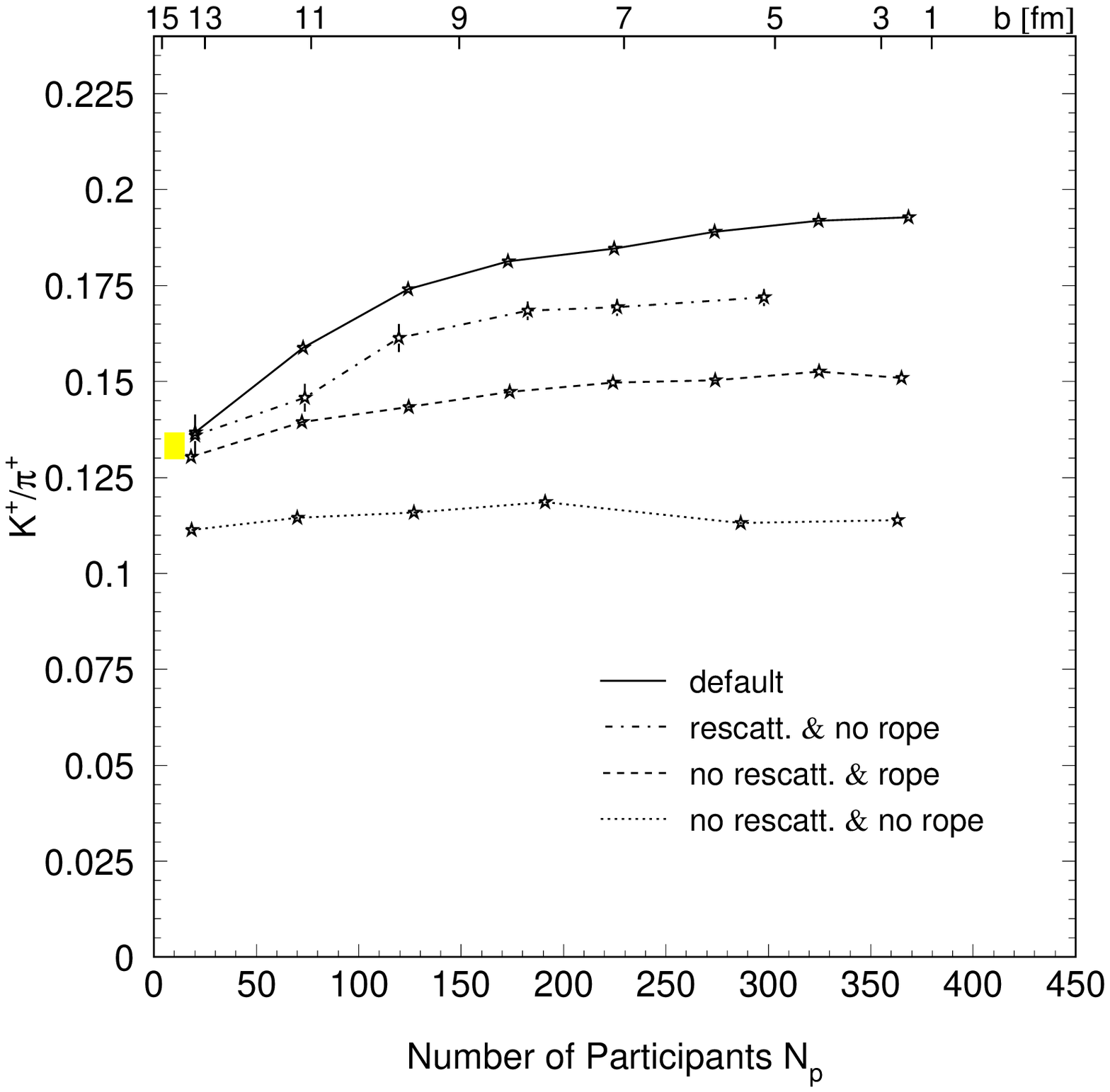}
\epsfxsize=0.5\textwidth\epsfbox[0 40 500 540]{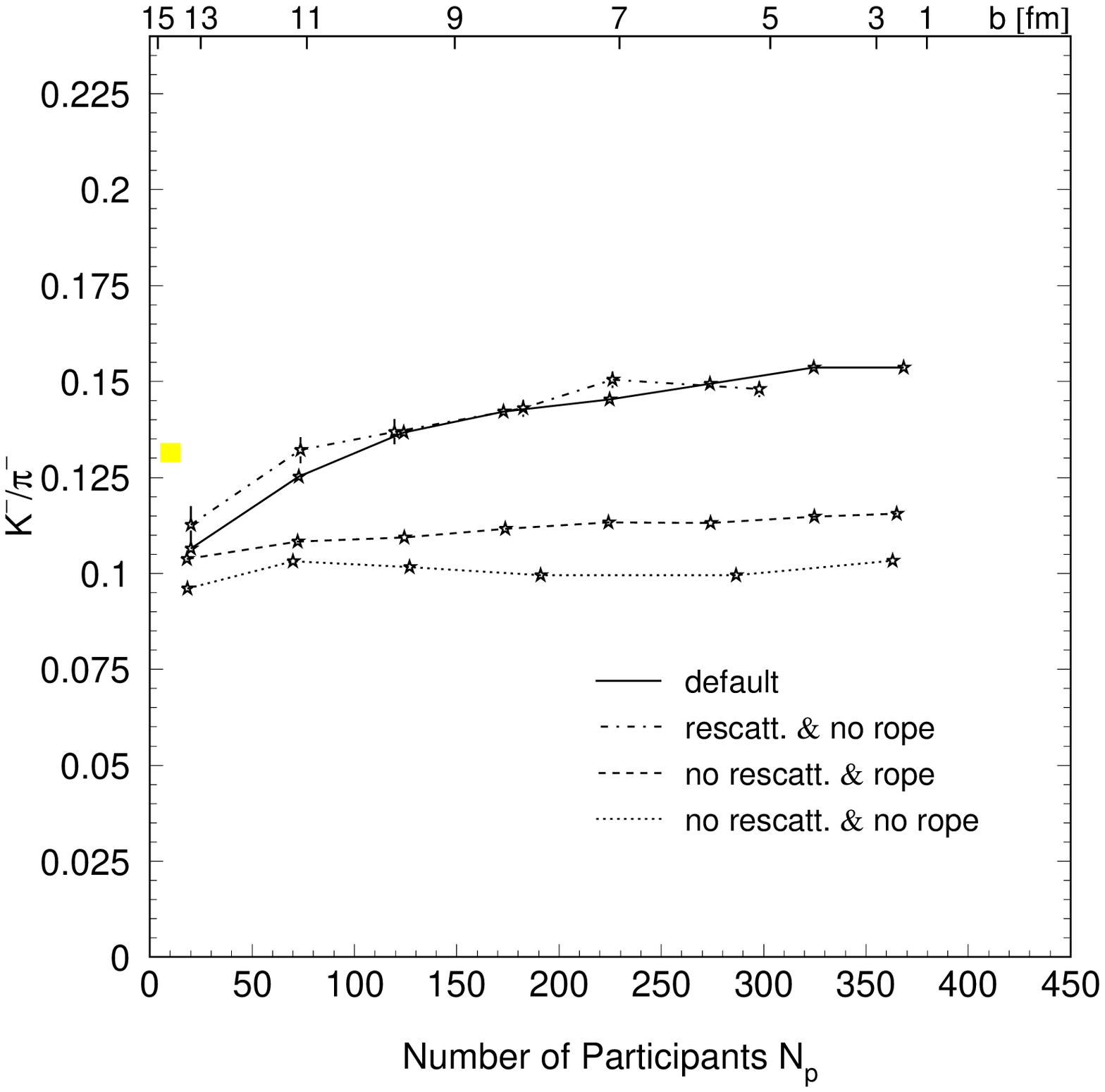}}
\caption{RQMD results of the kaon to pion total multiplicity ratios, 
$\kpip$ (left panel) and $\kpim$ (right panel), in Au+Au
collisions at RHIC energy as a function of 
the number of participants ($\np$). 
Shown are the RQMD results with meson-induced interactions and rope
formation both turned off (dotted line), with meson-induced interactions
turned off but rope formation included (dashed line), 
with meson-induced interactions included but rope formation turned off
(dot-dashed line), and with the default setting (solid line).
Errors are statistical only, either shown or smaller than the symbol size.
The shaded areas indicate the corresponding ratios in $p+p$ interactions
calculated by default RQMD.
The top axis indicates the average impact parameter ($b$) 
for Au+Au collisions at RHIC energy from default RQMD. 
The average impact parameters for 
the other settings 
have a systematic variation of $\pm 0.5$~fm from these values.}
\label{fig:npart_rhic}
\end{figure}

The dot-dashed curves in Figs.~\ref{fig:npart_ags_sps} and 
\ref{fig:npart_rhic} show the calculated $\kpi$ ratios in 
heavy-ion collisions at SPS and RHIC energies, respectively, by RQMD 
with rope formation turned off (and including particle rescattering).
The effect of rope formation is negligible at AGS energies.
The differences between these results and the default RQMD results
(solid curves) are small.
The small differences indicate that the effect of rope formation is 
small on $\kpi$ ratios once particle rescattering is considered.

It is interesting to investigate the case with both particle rescattering
and rope formation turned off, as this case corresponds to the simple
nucleon-nucleon collisions with only the complication of initial 
multiple scattering. The RQMD results for this simple case
are shown in the dotted curves in Figs.~\ref{fig:npart_ags_sps}
and \ref{fig:npart_rhic} for SPS and RHIC energies, respectively.
Comparing these results with the dashed curves 
(with rope and no rescattering) indicates that rope formation is able to 
make significant difference to $\kpip$ at both SPS and RHIC energies
if there is no subsequent rescattering. 
On the other hand, the effect on $\kpim$ is small.

Generally speaking, rope formation is an early phenomenon in RQMD
and particle rescattering takes place later.
Rope formation increases $\kpip$; particle rescattering washes
out these increases. Particle rescattering makes larger increases 
in $\kpip$ without rope formation (dot-dashed versus dotted curves)
than with rope formation (solid versus dashed curves). 
In other words, particle rescattering and rope formation do not act
on $\kpip$ additively. 
These observations are understandable if one considers rope formation 
a form of ``string/parton rescattering'' 
at early stages of heavy-ion collisions.

The $\kpim$ at SPS energy behaves differently from $\kpip$. 
Rope formation seems not to have an effect; 
the effect of particle rescattering is smaller than on $\kpip$. 
On the other hand, $\kpim$ at RHIC energy behaves similarly to $\kpip$.
This is expected because the net baryon density is low at RHIC energy.
If the net baryon density were zero, then $\kpip$ and $\kpim$ would be 
identical.

When both particle rescattering and rope formation are turned off 
as shown by the dotted lines in Figs.~\ref{fig:npart_ags_sps} and
\ref{fig:npart_rhic}, the $\kpi$ ratios are rather constant over $\np$.
Therefore, it is interesting to compare these results to $p+p$.
The calculated $p+p$ results are indicated by the shaded areas in 
Fig.~\ref{fig:npart_ags_sps} for AGS and SPS energies 
and in Fig.~\ref{fig:npart_rhic} for RHIC energy.
Note that $\kpip$ is slight lower in the isospin weighted
nucleon-nucleon ($N+N$) interactions than $p+p$, 
while $\kpim$ is slightly higher~\cite{Ahl98,Gaz91}.
As seen in the figures, the heavy-ion results with 
no rescattering and no rope are consistent with the $p+p$ values
(with two exceptions: the $\kpip$ at SPS and the $\kpim$ at RHIC energy).
The two exceptions are currently not understood. 
We remark that the $p+p$ results are consistent with extrapolations
of the measured heavy-ion data shown in Fig.~\ref{fig:npart_data}.

We now come back to the comparison between data and the RQMD results 
at SPS energy. None of the four curves for SPS energy shown in 
Fig.~\ref{fig:npart_ags_sps} seems to completely agree with the data.
We remark that the central collision data favor particle rescattering. 
In the following two sections, we only consider the default RQMD
which include both particle rescattering and rope formation.

\subsection{$\kpi$ excitation functions}

\begin{figure}
\centerline{
\epsfxsize=0.5\textwidth\epsfbox[0 100 520 550]{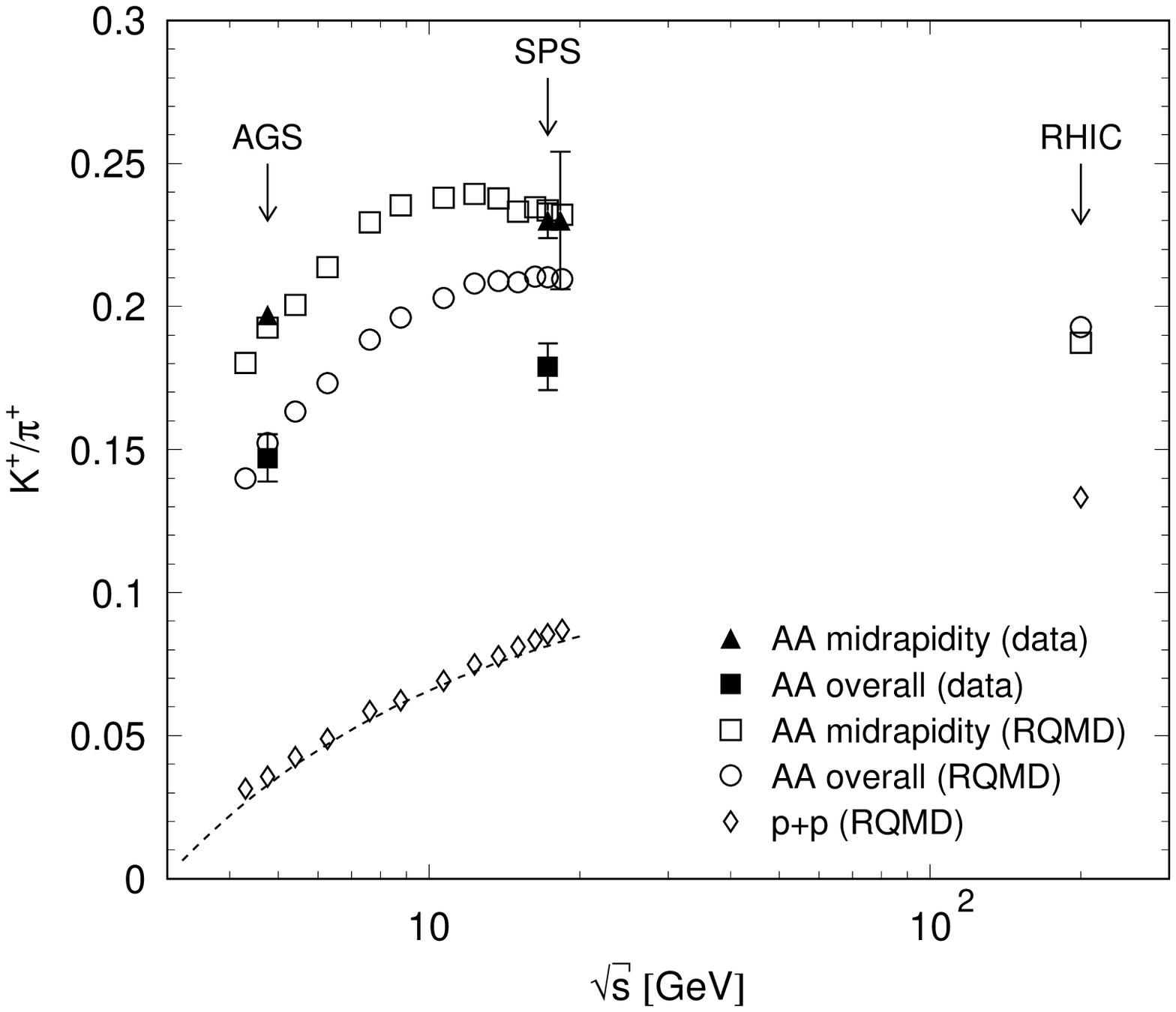}
\epsfxsize=0.5\textwidth\epsfbox[0 100 520 550]{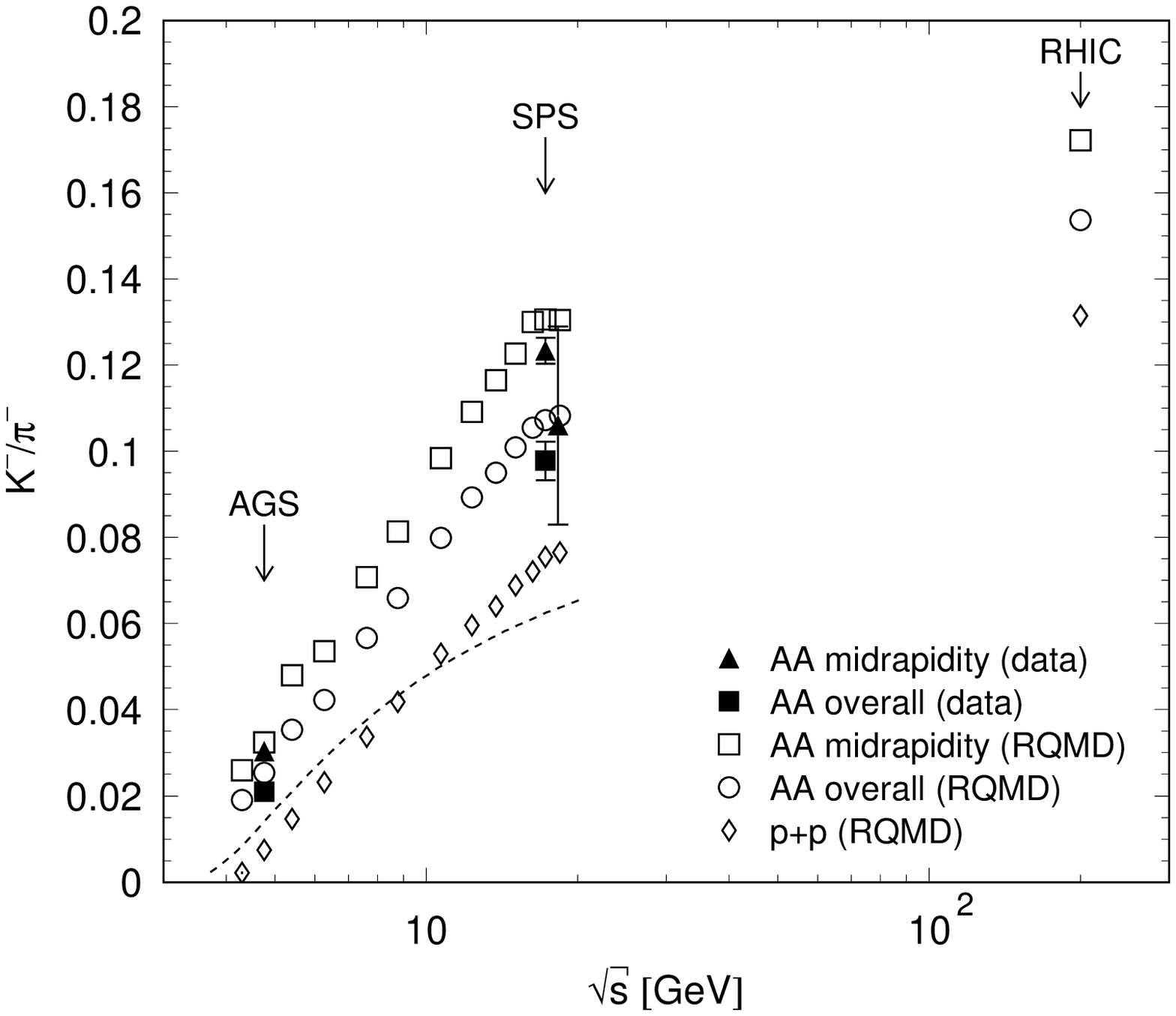}}
\caption{
RQMD results of the kaon to pion multiplicity ratios, 
$\kpip$ (left panel) and $\kpim$ (right panel), in Au+Au/Pb+Pb 
central collisions as a function of the nucleon-nucleon
center-of-mass energy ($\sqrt{s}$). 
The total multiplicity ratios are shown in open circles,
and the midrapidity ($1/3<y/y_{\rm beam}<2/3$)
multiplicity ratios are shown in open squares.
Those calculated for $p+p$ interactions are shown in open diamonds. 
The curves are parametrizations of experimental $p+p$ 
data~[\protect{\ref{Ros75:pp}}].
Data from AGS E866~[\protect{\ref{Wan96},\ref{Ahl98}}] and 
SPS NA44~[\protect{\ref{NA44}}] and NA49~[\protect{\ref{NA49}}]
are also shown for the total (filled square) and midrapidity 
(filled triangle) multiplicity ratios.
These data have been scaled up to approximately exclude decay contributions.
The NA44 data points (with the large error bars) are plotted 
at a slightly higher $\sqrt{s}$ for clarity.
Errors (either shown or smaller than the symbol size) 
are statistical only for both the RQMD results and the E866 data.
Errors on the NA44 data are combined statistical and systematic errors.
Errors on the NA49 data are quadratic sum of statistical and 
5\% systematic errors.}
\label{fig:sqrts}
\end{figure}

We study the excitation functions of $\kpi$ in the most central
Au+Au/Pb+Pb collisions ($\np > 350$) with RQMD, as a necessary 
theoretical baseline for the search of abnormal behavior.
The total multiplicity ratios are shown as open circles in 
Fig.~\ref{fig:sqrts} for $\kpip$ (left panel) and $\kpim$ (right panel).
The multiplicity ratios in the midrapidity region 
($1/3<y/y_{\rm beam}<2/3$) are shown as open squares.
The midrapidity ratios are larger than the total multiplicity ratios.

Data from AGS E866~\cite{Wan96,Ahl98} and SPS NA44~\cite{NA44} 
and NA49~\cite{NA49} are shown in the filled squares 
(total multiplicity ratios) and filled triangles (midrapidity ratios). 
In order to exclude the effect of decay contributions in the data,
the AGS $\kpip$ data are scaled up by 5\%, and the SPS $\kpip$ 
and $\kpim$ data are scaled up by 10\% 
The AGS data are well reproduced, and the SPS data are slightly 
overpredicted by RQMD. The overall agreement is good.

For comparison, the results for $p+p$ interactions from 
Fig.~\ref{fig:pp} are replotted in Fig.~\ref{fig:sqrts}.
The $\kpi$ ratios are significantly larger in heavy-ion collisions
than the same energy $p+p$ interactions.
Note that the isospin weighted $N+N$ interactions may be
better in comparison to heavy-ion collisions. 
However, comparing heavy-ion results to $p+p$ is more feasible 
because only $p+p$ data are experimentally available.

The $\kpip$ ratio in central heavy-ion collisions increases 
at low energy, partially due to the rapid increase in kaon 
production near threshold in $p+p$ interactions~\cite{Ros75:pp}. 
However, $\kpip$ saturates at high energies. 
Clearly, the continuous increase of $\kpip$ with $\sqrt{s}$
in $p+p$ is not seen in central heavy-ion collisions. 
This is a direct reflection of the amount of baryon stopping at 
different energies. The baryon midrapidity density decreases with 
$\sqrt{s}$~\cite{Wan98}, which is well described by the model. 
The ratio at midrapidity even decreases with $\sqrt{s}$ at high energies.
Maximum $\kpip$ appears at beam energy $\sim 40$~AGeV. 
This energy range is currently being studied at the SPS. 

As seen in Fig.~\ref{fig:sqrts} (right panel),
$\kpim$ continuously increases with $\sqrt{s}$. 
The dependence is distinctly different from that of $\kpip$.
The ratio is larger in heavy-ion collisions than $p+p$ interactions at 
all corresponding energies, but follow the general trend seen in $p+p$.
The increase in $\kpim$ from $p+p$ to heavy-ion collisions at the same 
energy is significantly smaller than that in $\kpip$.
The statement can be made stronger when taken into account that 
the charge asymmetry results in a higher $\kpip$ and 
a lower $\kpim$ in $p+p$ than the isospin weighted $N+N$ interactions.

At RHIC energy, respectively for $\kpip$ and $\kpim$,
RQMD predicts a total multiplicity ratio of $0.19$ and $0.15$,
and a midrapidity multiplicity ratio of $0.19$ and $0.17$.
The predictions provide a baseline for comparison to 
experimental data which are expected soon.

\begin{figure}
\centerline{
\epsfxsize=0.6\textwidth\epsfbox[0 100 520 550]{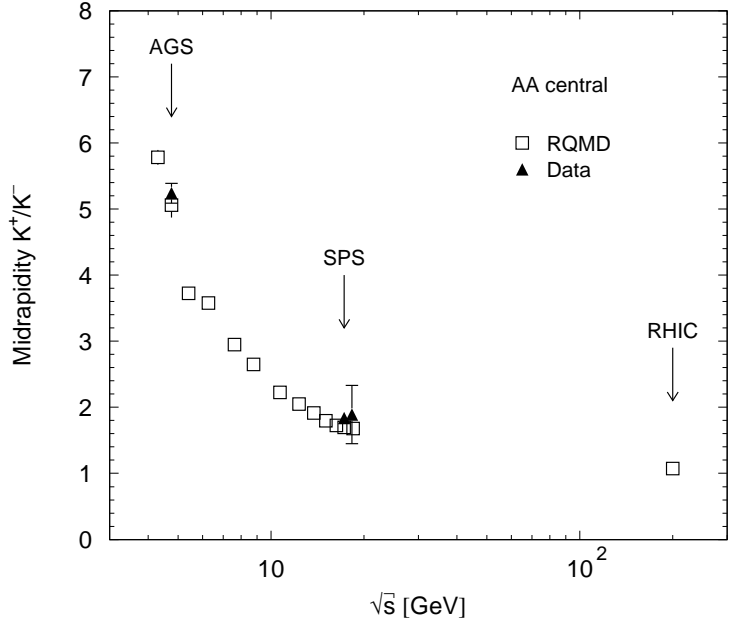}}
\caption{
The midrapidity ($1/3<y/y_{\rm beam}<2/3$) kaon multiplicity ratio 
$\kpkm$ in Au+Au/Pb+Pb central collisions as a function of the 
nucleon-nucleon center-of-mass energy ($\sqrt{s}$). 
The RQMD results are shown in open squares.
Data from AGS E866~[\protect{\ref{Ahl98}}] and preliminary data from
SPS NA44~[\protect{\ref{NA44}}] and NA49~[\protect{\ref{NA49}}] 
are shown in filled triangles.
The NA44 data point (with the large error bar) is plotted at a slightly
higher $\sqrt{s}$ for clarity.
Except for the NA44 point, errors are statistical only, 
either shown or smaller than the symbol size.
Error on the NA44 data point is combined statistical 
and systematic error.}
\label{fig:kpkm}
\end{figure}

The distinct difference between the $\kpip$ and $\kpim$ 
excitation functions results from the different production 
mechanisms for $K^+$ and $K^-$, which are connected to 
the presence of the baryon density~\cite{Sor95:mass}.
The effect of the baryon density can be made more clearly 
in the kaon multiplicity ratio ($\kpkm$) versus $\sqrt{s}$. 
Figure~\ref{fig:kpkm} shows the midrapidity ratio 
of $\kpkm$ as a function of $\sqrt{s}$.
The ratio decreases steadily with $\sqrt{s}$. 
For comparison, the E866~\cite{Ahl98}, NA44~\cite{NA44}, 
and NA49 data~\cite{NA49,Wan98} are also shown. 
The data are well reproduced by RQMD.

\subsection{$\kpi$ enhancement}

Now let us turn to the enhancement in $\kpi$ ratios.
We define the enhancement factors as the ratios of $\kpip$
and $\kpim$ in central heavy-ion collisions over those
in $p+p$ interactions at the corresponding energy.
Note that this is not a perfect definition.
One may argue that a division by the isospin weighted $N+N$ results 
is a better definition, but it is not experimentally preferred.

\begin{figure}
\centerline{
\epsfxsize=0.53\textwidth\epsfbox[0 100 520 550]{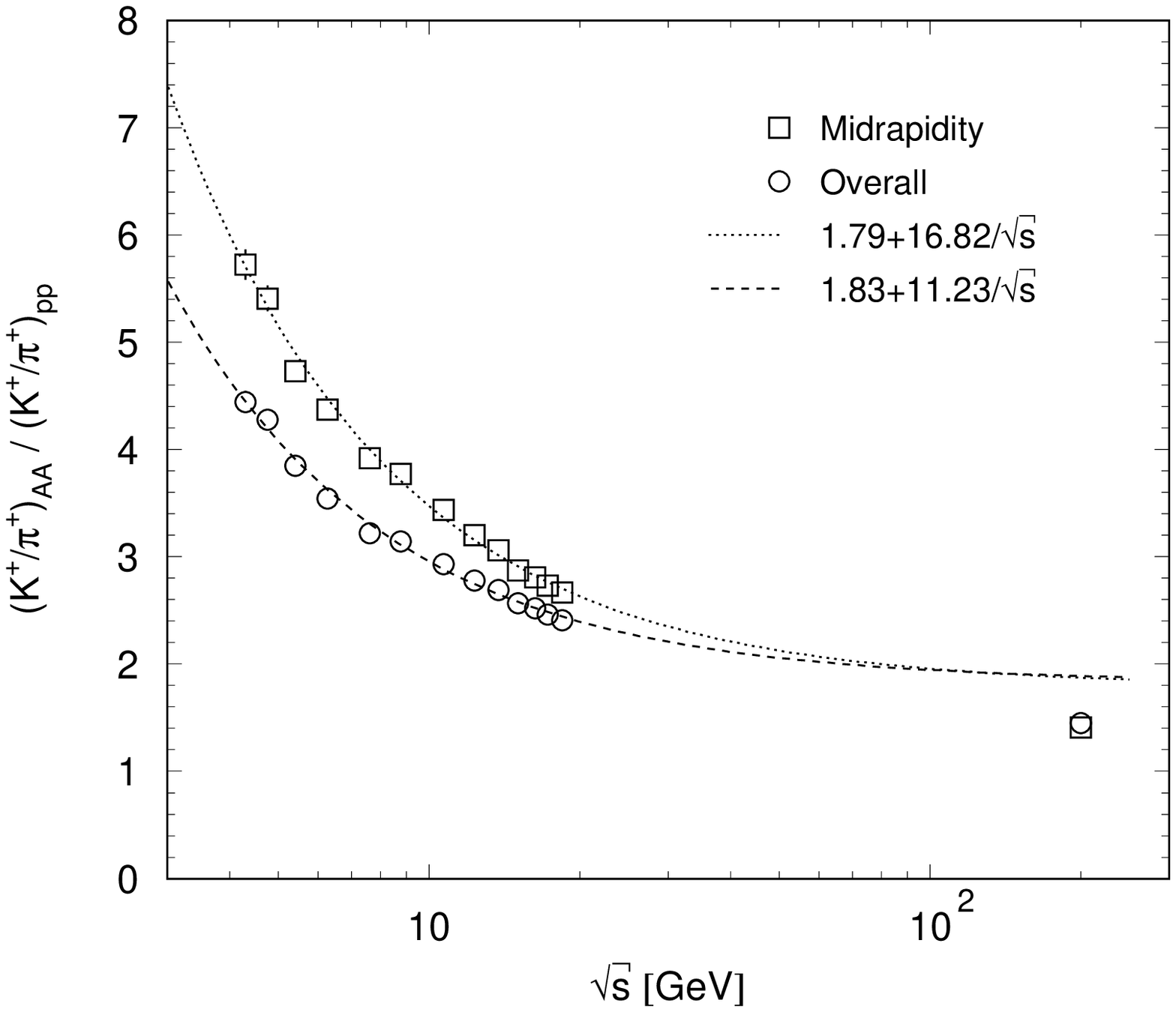}
\epsfxsize=0.53\textwidth\epsfbox[0 100 520 550]{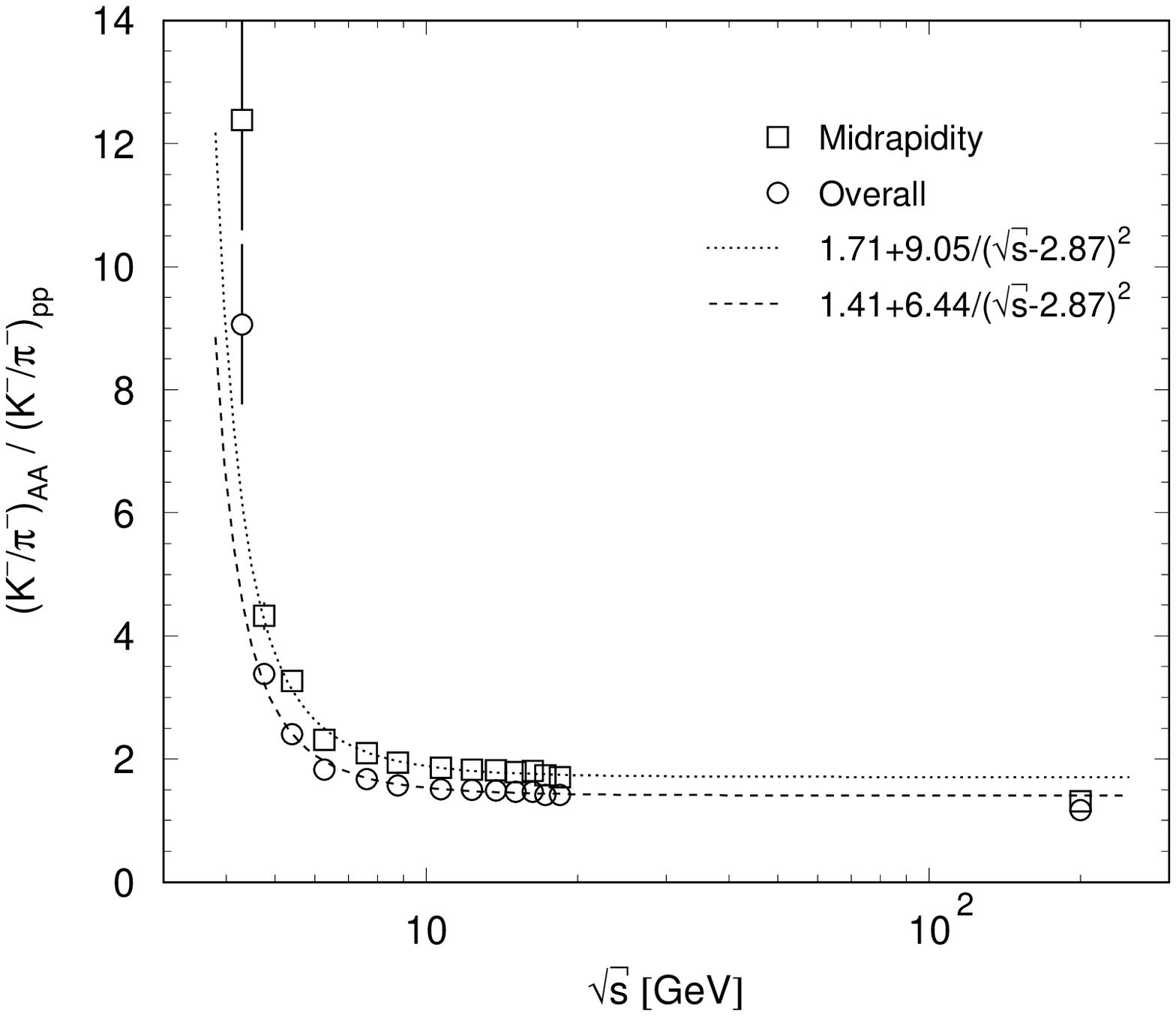}}
\caption{
RQMD results of the strangeness enhancement factor, the ratios of 
$\kpip$ (left panel) and $\kpim$ (right panel) in central heavy-ion 
collisions over those for the same energy $p+p$ interactions, 
as a function of the nucleon-nucleon center-of-mass energy ($\sqrt{s}$). 
The enhancement in the total multiplicity ratios are shown in open circles;
the enhancement in the midrapidity ($1/3<y/y_{\rm beam}<2/3$) 
multiplicity ratios are shown in open squares.
Errors are statistical only, either shown or smaller than the symbol size.
The curves are two-parameter fits to the results 
excluding those at RHIC energy.}
\label{fig:enhancement}
\end{figure}

Figure~\ref{fig:enhancement} shows the results of the 
enhancement in $\kpip$ (left panel) and $\kpim$ (right panel).
The enhancement is larger at low energy for both $\kpip$ and $\kpim$.
This is partially due to the effect of kaon production threshold.
As discussed before, meson-baryon interactions involving resonances 
are the primary processes for $\kpi$ enhancement at AGS energy.
At low energies near threshold, particle rescattering is most prominent.
Take an extreme case that the beam energy is below the threshold:
No kaons can be produced in $p+p$ interactions; 
however, kaons can be produced in heavy-ion collisions 
by invoking the multi-step processes where more than 
two particles cooperate to overcome the production threshold. 
In this extreme case, the enhancement is infinity.

The enhancement in $\kpip$ drops more quickly at lower $\sqrt{s}$.
Over the whole range of $\sqrt{s}$ studied between AGS and SPS 
energies, the enhancement factor decreases smoothly with $\sqrt{s}$.
The quick drop at low $\sqrt{s}$ is partially due to the threshold 
effect mentioned above. 
The shallow tail of the enhancement factor at large $\sqrt{s}$
results from the flattening of $\kpip$ at high $\sqrt{s}$ in 
heavy-ion collisions and the continuous increase of the ratio 
with $\sqrt{s}$ in $p+p$ interactions.

As shown in Fig.~\ref{fig:enhancement} (right panel), 
the enhancement in $\kpim$ drops very quickly at low $\sqrt{s}$.
The behavior is consistent with the high $K^-$ production threshold.
However, at large $\sqrt{s}$ the enhancement factor is almost constant. 

Since the enhancement factors of both $\kpip$ and $\kpim$ have a weak 
dependence on $\sqrt{s}$ at high energies, it is possible, 
without RQMD calculations for RHIC energy, 
to predict the enhancement factors at RHIC energy by extrapolation.
In order to do so, we parametrize the enhancement factor as 
a function of $\sqrt{s}$. 
We choose to parametrize the $\kpip$ enhancement factor
by the functional form $A+B/\sqrt{s}$.
Fits to the enhancement factors shown in the left panel of 
Fig.~\ref{fig:enhancement} (of course excluding that at RHIC energy)
yield $1.8+11.2{\rm GeV}/\sqrt{s}$ and $1.8+16.8{\rm GeV}/\sqrt{s}$,
respectively for the enhancement in the overall $\kpip$ 
and the midrapidity $\kpip$.
The $\chi^2/N_{DF}$ of the fits are on the order of 1.5.
The fit results are superimposed in Fig.~\ref{fig:enhancement} 
(left panel) as the dashed and dotted curves. 
The two curves essentially overlap at RHIC energy, 
and slightly miss the calculated results.

Because of the sharp drop of the $\kpim$ enhancement factor
at low energy, we choose to parametrize the $\kpim$ enhancement 
factor by the functional form $A+B/(\sqrt{s}-2.87{\rm GeV})^2$, 
where 2.87~GeV is the kaon pair production threshold in  
$NN\rightarrow NNK^+K^-$.
Fits to the enhancement factor in the right panel of
Fig.~\ref{fig:enhancement} (excluding that at RHIC energy) yield 
$1.4+6.4{\rm GeV}/(\sqrt{s}-2.87{\rm GeV})^2$ and
$1.7+9.1{\rm GeV}/(\sqrt{s}-2.87{\rm GeV})^2$,
respectively for the enhancement in the overall $\kpim$ 
and the midrapidity $\kpim$. 
The $\chi^2/N_{DF}$ of the fits are on the order of 2.5.
The large $\chi^2/N_{DF}$ is mainly due to the lowest 
$\sqrt{s}$ point.
If the lowest $\sqrt{s}$ point is excluded from the
fits, $\chi^2/N_{DF}$ becomes about 1.5 without 
essential change in the fit results.
The fit results are superimposed in Fig.~\ref{fig:enhancement} 
(right panel) as the dashed and dotted curves. 
They can satisfactorily describe the enhancement factor at 
RHIC energy as directly calculated by RQMD.

\section{Conclusions}\label{conclusion}

We have reported a systematic study of the kaon to pion multiplicity
ratios, $\kpip$ and $\kpim$, in heavy-ion collisions as a function of 
the bombarding energy from AGS to RHIC energy, using the RQMD model.
We have demonstrated that the kaon and pion multiplicity data in $p+p$
interactions are well reproduced by RQMD, 
validating the comparison between data and RQMD for heavy-ion collisions.

The RQMD-calculated ratios in heavy-ion collisions are higher than 
those in $p+p$ interactions at the same energy, and increase from peripheral
to central collisions.
By comparing the results to those calculated by RQMD with particle 
rescattering (meson-induced interactions) turned off, we conclude 
that the $\kpi$ enhancement in central collisions with respect to 
peripheral collisions and $p+p$ interactions is mainly due to 
meson-induced interactions, especially at the low AGS energy.
We further conclude that rope formation does not change the $\kpi$
ratios significantly once particle rescattering is considered.

It is found that the $\kpi$ enhancement in central heavy-ion 
collisions over $p+p$ interactions is larger at AGS than 
SPS energy, and decreases smoothly with bombarding energy.
This behavior is consistent with the combination of the threshold 
effect of kaon production in $p+p$ interactions and the dropping 
baryon density in heavy-ion collisions with increasing bombarding energy.

The RQMD model reasonably describes the available 
$\kpip$ and $\kpim$ data in heavy-ion collisions.
The observed $\kpi$ enhancement at the AGS and SPS can be understood
in the RQMD model with hadronic rescattering 
and string degrees of freedom.

The RQMD $\kpi$ results agree with experimental data better 
at midrapidity in the most central collisions. 
This is not surprising since it is the case where equilibrium 
is most likely to be reached.
There the details of the description in any model is less relevant.

The total multiplicity ratios at RHIC energy are 
predicted by RQMD to be $\kpip=0.19$ and $\kpim=0.15$.
The midrapidity ratios at RHIC energy are 
predicted by RQMD to be $\kpip=0.19$ and $\kpim=0.17$.

\section*{Acknowledgments}

We are grateful to Dr. G. Odyniec, Dr. G. Rai, and Dr. H.G. Ritter 
for valuable discussions.
This work was supported by the U.~S.~Department of Energy under Contracts
No. DE-AC03-76SF00098, DE-FG02-88ER40388, and DE-FG02-89ER40531, and used 
resources of the National Energy Research Scientific Computing Center.

\onecolumn

\newpage

\vspace{-9.5mm}

\end{document}